\documentclass[twocolumn,twoside,slac_two]{revtex4}
\usepackage{graphicx}
\usepackage{fancyhdr}
\newcommand{\chandra}{{\it Chandra}}
\newcommand{\hst}{{\it HST}}
\newcommand{\rosat}{{\it ROSAT}}
\newcommand{\xmm}{{\it XMM-Newton}}

\newcommand{\lum}{\thinspace\hbox{$\hbox{ergs}\thinspace\hbox{s}^{-1}$}}
\newcommand{\sss}{M101 ULX-1}
\def\ref{\par\noindent\hangindent 20pt}

\def\spose#1{\hbox to 0pt{#1\hss}}
\def\laeq{\mathrel{\spose{\lower 3pt\hbox{$\mathchar"218$}}
     \raise 2.0pt\hbox{$\mathchar"13C$}}}
\def\gaeq{\mathrel{\spose{\lower 3pt\hbox{$\mathchar"218$}}
     \raise 2.0pt\hbox{$\mathchar"13E$}}}

\begin{document}

\title{Multiwavelength Observations and State Transitions of an
  Ultra-luminous Supersoft X-ray Source: Evidence for an
  Intermediate-Mass Black Hole}

%

\author{A.K.H. Kong}
\affiliation{Harvard-Smithsonian Center for Astrophysics, 60 Garden 
Street, Cambridge, MA 02138, USA}
\affiliation{MIT Center for Space Research, 77 Massachusetts Avenue,
  Cambridge, MA 02139, USA}

\author{M.P. Rupen, L.O. Sjouwerman}
\affiliation{National Radio Astronomy Observatory, Socorro, NM 87801, USA}

\author{R. Di\,Stefano}
\affiliation{Harvard-Smithsonian Center for Astrophysics, 60 Garden 
Street, Cambridge, MA 02138, USA}

\begin{abstract}

We report the results of \chandra\ and \xmm\ observations of an
ultra-luminous supersoft X-ray source in M101. M101 ULX-1
underwent 2 outbursts in 2004 during which the peak bolometric
luminosities reached $10^{41}$\lum. The outburst spectra were very soft and can generally be fitted with a
blackbody model with temperatures of 50--160 eV. In two of
the observations, absorption edges at 0.33 keV, 0.56 keV, 0.66 keV,
and 0.88 keV were found. A cool accretion disk was also found in the
2004 December outburst. During the low luminosity state, a power-law
tail was seen up to 7 keV. It is clear the source changed from a
low/hard state to a high/soft state. In addition, it showed at least 5
outbursts between 1996 and 2004. This is the first ultra-luminous X-ray
source for which recurrent outbursts with state transitions similar to
Galactic X-ray binaries have been observed. From the {\it Hubble Space
  Telescope} data, we found an optical counterpart to the
source. During the 2004 outbursts, we also performed radio and ground-based
optical observations. All the results strongly suggest that the
accreting object is a $> 2800 M_\odot$ black hole.

\end{abstract}

\maketitle

\thispagestyle{fancy}


\section{Introduction}

Recent high angular resolution X-ray observations reveal that there are a
large number of ultra-luminous X-ray sources (ULXs) in many nearby
galaxies. ULXs are luminous ($L_X > 10^{39}$
\lum) non-nuclear X-ray point sources with apparent X-ray luminosities
above the Eddington limit for a $\sim 10 M_{\odot}$ black hole (BH). While some
ULXs have been associated with supernovae, many are thought
to be accreting objects with X-ray flux variability observed on
timescales of hours to years.  A
natural possibility is that the compact object is an intermediate-mass 
black hole (IMBH) with mass of
$\sim 10^{2-4} M_{\odot}$ \cite{mc}. The origin of
such objects remains uncertain. Some ULXs may
have a stellar-mass black hole with beamed emission 
\cite{king01,kording02}.
While current observations are inconclusive about the
nature of ULXs, recent X-ray observations show that some ULXs have a
cool accretion disk ($kT\sim 0.1$ keV), suggesting the presence of IMBHs
\cite{miller03,miller04}.

While the majority of ULXs have X-ray emission from 0.1 to 10 keV, a few 
ULXs
have very soft spectra with no X-ray emission above 1 keV 
\cite{fab03,kd03,dk03}, similar to
supersoft X-ray sources (SSSs) in the Milky Way. The high luminosities
of ultra-luminous SSSs are inconsistent
with typical nuclear burning white dwarf models for Galactic SSSs. These
ultra-luminous SSSs could, however, very
well be IMBHs. Their
luminosities and temperatures are consistent with what is predicted for
accreting BHs with masses between roughly 100 and 1000 $M_{\odot}$.
Alternatively, outflows from stellar-mass BHs could also
achieve such a high luminosity and low temperature \cite{kp03}. 

In this paper, we report a series of X-ray/optical/radio observations of 
the ultra-luminous SSS in M101 (CXOU J140332.3+542103; M101 ULX-1 hereafter) 
during the low states and outbursts.

\section{X-ray Observations}

\begin{figure*}[t]
\includegraphics[width=3.1in]{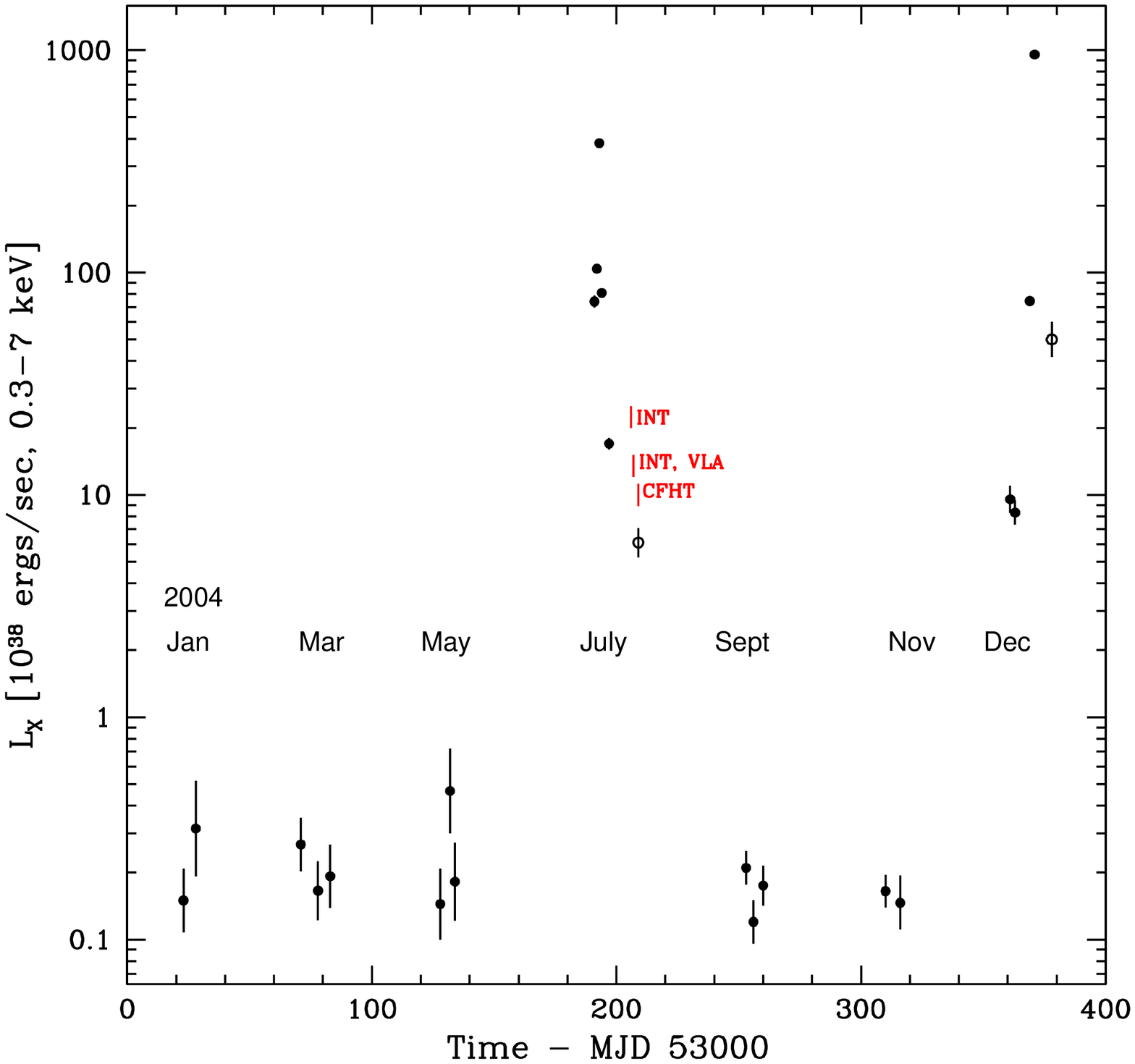}
\includegraphics[width=3.1in]{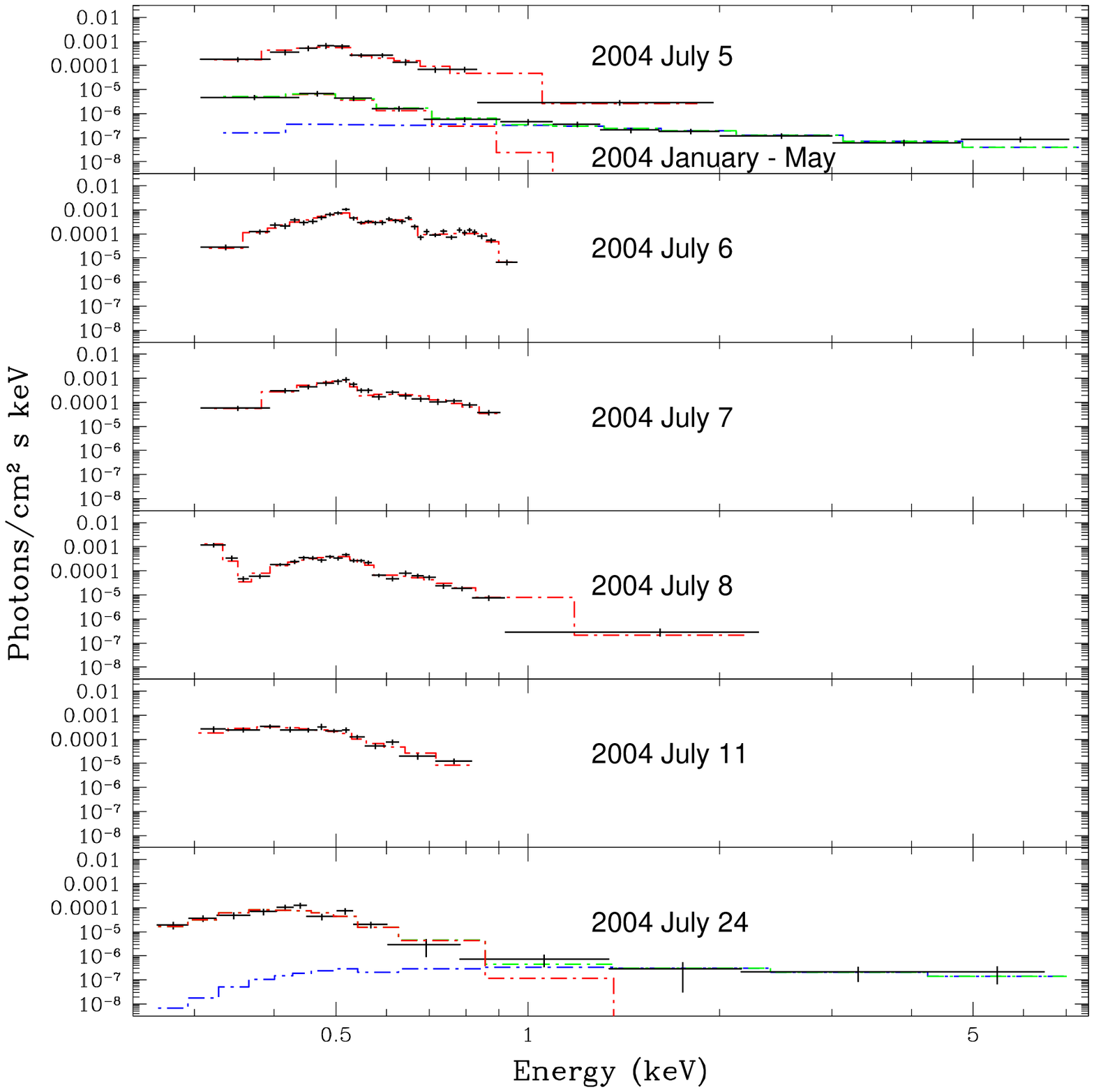}
\caption{Left: The light curve of M101 ULX-1 in 2004 as 
observed by \chandra\ (solid circles) and \xmm\ (open circles). It is very 
clear that the source stays 
at $\sim 10^{37}$ ergs s$^{-1}$ during the low luminosity state. When it 
went into an 
outburst in 2004 July and December, the peak 0.3-7 keV luminosity was near 
$10^{41}$\lum, a factor of $\sim1000$ comparing to the low state. We also 
show 
the optical (INT and CFHT) and radio (VLA) coverage in the figure. It is worth 
noting that the CFHT observation was taken simultaneously with \xmm. 
Right: Unfolded spectra of M101 ULX-1. The total spectrum, 
blackbody component, and power-law component are shown in green, red, and 
blue, respectively. During the 2004 July outburst, the spectra can be 
fitted with blackbody/disk blackbody model with temperatures of 50-100 eV. 
In two of the observations, absorption edges at 0.33, 0.57 (July 8), 0.66, 
and 0.88 keV (July 6) were found. On July 7, the peak  bolometric 
luminosity reached $10^{41}$\lum\ (which is one of the most luminous 
ULXs). The \xmm\  spectrum taken on July 23 can be fitted with a blackbody 
($kT=53$ eV) and a power-law ($\alpha=0.72$) model. The combined Jan-May 
spectrum is similar to the \xmm\ spectrum, but 
with much lower luminosities ($\sim 10^{37}$\lum). 
} 
\end{figure*}

M101 ULX-1 is one of the most luminous ULXs. It was discovered with
{\it ROSAT} and was
confirmed as a SSS with a blackbody temperature of about 100 eV, with {\it
Chandra} \cite{pence01,mukai03,dk03}.
During 2000 March, {\it Chandra} detected it at
$L_X\sim4\times10^{39}$ ergs s$^{-1}$, and then in 2000 October, its
luminosity was around $10^{39}$ ergs s$^{-1}$. 
In 2004, {\it Chandra} conducted
a monitoring program for M101. Figure 1 (Left) shows
the long-term X-ray lightcurve of M101 ULX-1 from 2004 January to 
2005 January.
M101 ULX-1 was near the
detection limit during January, March, and May; the X-ray spectra were
harder with a power-law shape (see Figure 1 Right), and the X-ray 
luminosity was
about $2\times10^{37}$ ergs s$^{-1}$,
a factor of about $10^2$ fainter than
that during the outbursts in 2000 \cite{kong04}.
The source was found to be in
outburst during the July 5 observation, with an X-ray luminosity of
about $7\times10^{39}$ ergs s$^{-1}$. Data taken on July 6, 7,
and 8 show that the source was in outburst with a peak
bolometric luminosity (for assumed isotropic emission)
of about $10^{41}$ ergs s$^{-1}$ \cite{kong04}. In general,
the X-ray spectra are best described with an absorbed blackbody model
with temperatures of $\sim50-100$ eV (see Figure 1). In addition, we found
absorption
edges at 0.33, 0.57, 0.66, and 0.88 keV in two of the high state spectra.
These features may signal the presence of highly ionized gas in the
vicinity of the accretor (e.g., warm absorber). A DDT \xmm\ observation
was made on July 23 and the luminosity was about
$6\times10^{38}$ ergs s$^{-1}$. A harder X-ray spectrum with a
power-law tail ($kT=53$ eV and $\alpha=0.72$) was seen up to 7 keV. More 
recently, {\it Chandra}
observations made in September and November indicated that the source
returned to the
low state with a power-law spectrum and a luminosity of $\sim 
2\times10^{37}$ ergs s$^{-1}$. In 2004
December, the source was in outburst again with very soft spectra
($kT=50-160$ eV) and a peak luminosity of about
$10^{41}$ ergs s$^{-1}$. During the rise of the outburst, the spectra 
were supersoft with blackbody temperatures of 40--70 eV. The source 
showed a cool accretion disk near the peak
of the outburst on 2005 January 1. The X-ray spectrum can be fitted with a 
disk blackbody model with $kT_{DBB}=0.16$ keV which is significantly 
harder than previous supersoft X-ray spectra. The peak luminosity of the 
2004 December outburst is about $9\times10^{40}$\lum\ which is very 
similar to the 2004 July outburst. A DDT \xmm\ observation made on 2005 
January 8 indicated that the source was in the decline stage 
($L_X=6\times10^{39}$\lum) and the 
X-ray spectrum returned to a supersoft state ($kT=56$ eV) \cite{kong05}. 

\section{Optical Observations}

M101 ULX-1 is located near star forming regions in a spiral arm. Figure 2 
shows a far UV (1350--1750 Angstroms) image of M101 taken with
GALEX. While GALEX cannot detect M101 ULX-1 because of the sensitivity
and confusion, 
the ultra-luminous SSS is about $10''$ away from a bright UV source. 
Based on the position provided by \chandra, we searched for optical
counterpart of M101 ULX-1 in archival {\it Hubble Space Telescope (HST)} data. The region of
M101 ULX-1 was observed with \hst\ using Wide Field Planetary Camera 2
(WFPC2) and Advanced Camera for Surveys (ACS) in 1994, 1995, and 2002. After reducing the data
with standard procedures and correcting the
astrometry of the \hst\ and \chandra\ images, we 
found a blue object 
($V=23.8$, $B-V=-0.2$) within the $0.6''$ \chandra\ error circle
(Figure 3). At the distance of M101 (d=6.7 Mpc), the absolute
magnitude corresponds to $M_V=-5.3$. The magnitudes and colors of the
blue star are consistent with an OB star. During the 2004 July outburst, 
we performed a series of target-of-opportunity observations at
ground-based optical telescopes 
(see Figure 1 and 3). These observations included WIYN 3.5m, the 4.2m William 
Herschel Telescope (WHT), the 2.5m Isaac Newton
Telescope (INT), and the 3.6m Canada-France-Hawaii Telescope (CFHT). 
Unfortunately, WIYN and WHT observations were affected by bad weather and 
bright moon. M101 ULX-1 was barely detected in the INT images but was 
clear seen in the CFHT images even under very bad 
seeing (Figure 3). It is worth noting that the CFHT observations were 
performed {\it simultaneously} with \xmm. These are the {\it first} 
simultaneous X-ray/optical observations for an ULX.
During the 2004 December outburst, we also performed an optical follow-up 
observation with the WIYN 3.5m on 2005 January 17 and M101 ULX-1 was 
clearly detected.

\begin{figure}[t]
\centering
\includegraphics[width=3.1in]{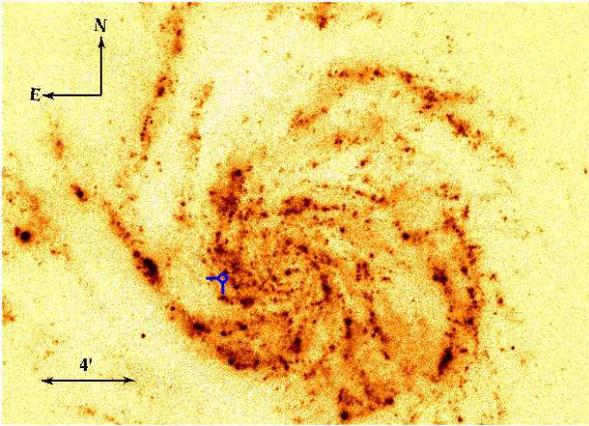}
\caption{GALEX far UV (1350--1750 Angstroms) image of M101. The blue 
circle is the location of M101 ULX-1.}
\end{figure}

\begin{figure*}
\centering
\includegraphics[width=6in]{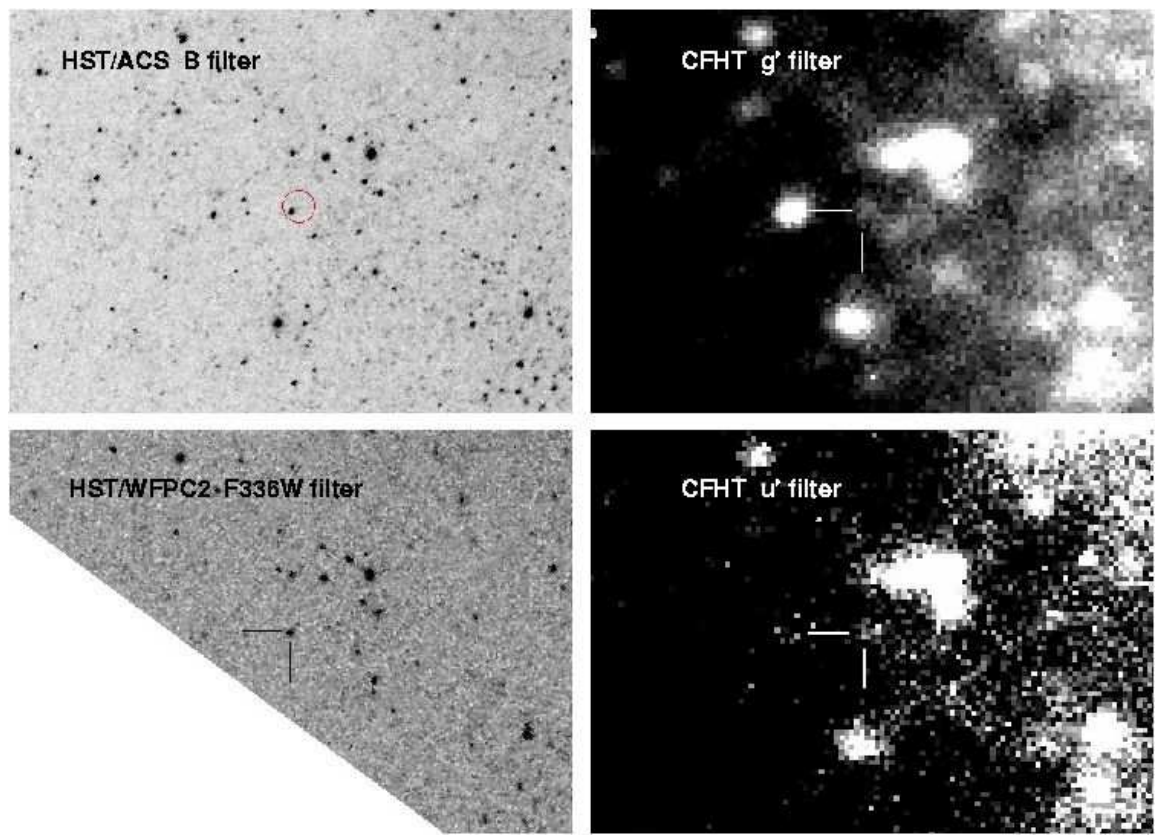}
\caption{Upper Left: \hst/ACS B band image (900s) of the field near the 
SSS. The image has been corrected for astrometry using 2MASS catalog. The 
$0.6''$ \chandra\ error circle is indicated. The $B$ magnitude is about 
23.6. Lower Left: HST/WFPC2 $U$ band (F336W) image (1200s) of the same 
field. The 
$U$ magnitude of the source is about 22.3. Upper Right: CFHT $g'$ band 
image (600s) of the same field. The image was taken simultaneously with 
\xmm\ on 2004 July 23. Lower Right: CFHT $u'$ band image (600s) of the 
same field. 
} 
\end{figure*}

\section{Radio Observations}

After the discovery of the outburst of M101 ULX-1 in 2004 July with 
\chandra, we observed M101 ULX-1 under target-of-opportunity time with 
Very Large Array 
(VLA) D-array at 4.86 (14
arcsec resolution) and 8.46 GHz (8.4 arcsec resolution; Fig. 4) for 
about 1.5 hours on 2004 July 21. There is no radio counterpart with a
$3\sigma$ detection limit of 0.075 mJy/beam at 8.46 GHz. We also
searched for the radio emission from archival VLA data. In the 2002
May 7 A-array 1.4 GHz (1.5 arcsec resolution) data, the source was 
not detected in the 6-hour observation but there is a $2\sigma$ hint of a
source about $1''$ away from M101 ULX-1; the 3$\sigma$ detection limit of the 
observation is 0.075 mJy/beam.

\begin{figure*}[t]
\includegraphics[width=6in]{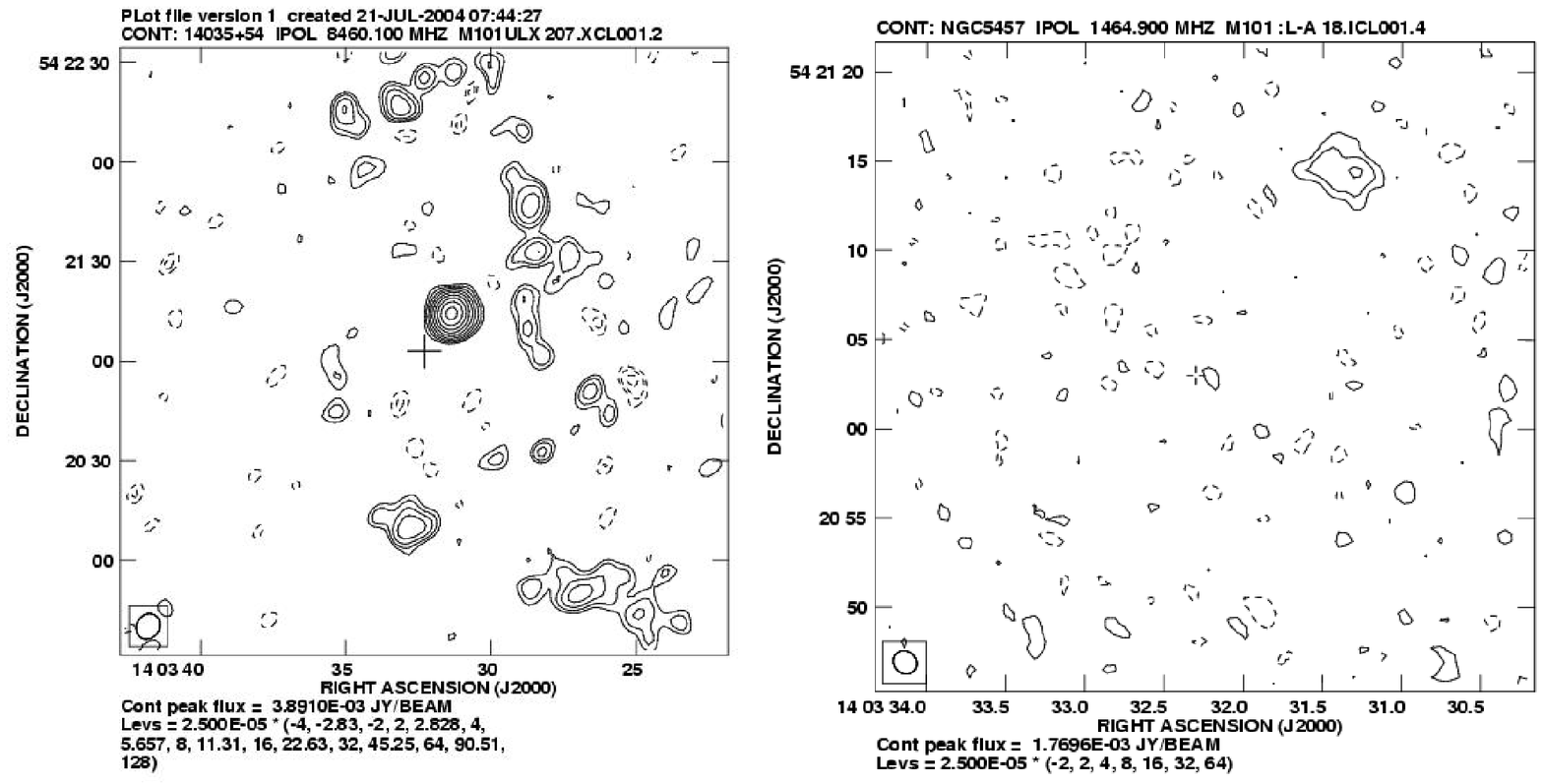}
\caption{VLA observations of M101 ULX-1 with 4.86 GHz (14 arcsec 
resolution; Left) and 1.4 GHz (1.5 arcsec resolution; Right). The cross is 
the \chandra\ position of M101 ULX-1.
}
\end{figure*}

\section{Discussion}

M101 ULX-1 is a very unique ULX. It has a very soft blackbody spectrum 
during outburst while it has a hard power-law tail in the low state. It 
also has recurrent outbursts which provide an excellent opportunity
to explore a totally new variability timescale for an ULX. Only a handful 
of transient ULXs have been detected. This may be due to the lack of 
repeated observations of
nearby galaxies. M101 ULX-1 is the only ULX that has been observed
regularly with X-ray telescopes. Hence, rapid multiwavelength follow-up 
observations become possible and meaningful. The X-ray/optical/radio 
observaitons of M101 ULX-1 reported here is the first and only rapid 
follow-up observations for an ULX.
   
Although over $100$ ULXs have been discovered \cite{mc}, 
only a handful of sources have $L_X\gaeq 10^{41}$ ergs s$^{-1}$, assuming 
isotropic radiation \cite{mat01,gao03,dm04,sm04}.
A luminosity this high is difficult to achieve in
an X-ray binary unless the accretor has a mass greater than
roughly $10\, M_\odot.$ We do have evidence that \sss\
is likely to be an X-ray binary, since its luminosity has been
observed to change by a factor of $\gaeq 10$ on a time scale of hours 
\cite{mukai03,dk03}.
While it is possible that our luminosity estimates are
higher than the true luminosity, the effects that lead to
overestimates, such as beaming, or various anisotropies,
tend to change estimated luminosities downward by a factor of roughly
$10.$ Its unusually high bolometric luminosity ($10^{41}$ ergs 
s$^{-1}$),
coupled with its
short-time-scale time variability, therefore make \sss\
a good candidate for an accreting IMBH.  For instance, anisotropic
X-ray emission can result super-Eddington luminosity for a
stellar-mass BH \cite{king01}. However, in order to achieve such a high 
luminosity
for a stellar-mass BH,
extreme beaming is required and the disk is expected to be
much hotter. Similarly, the temperature of radiation pressure-dominated 
accretion disk
model proposed is too high \cite{bel02}.
The pure blackbody spectrum
also makes relativistically beaming unlikely \cite{kording02}.

Its soft spectrum in the high state is another important piece of 
evidence.
An accretion disk around a very massive
BH is expected to produce supersoft X-ray emission \cite{di04}.
If we used the 90\% lower
limits of the inner disk temperature derived from the disk blackbody fits,
the BH mass is estimated to be $> 2800 M_\odot$ \cite{mak00}.
This is consistent with
prior work on IMBH models for
ULXs \cite{mc}. It also complements other
work on evidence for cool disks
in ULXs that has also been considered as evidence for
IMBH models \cite{miller03,miller04} although a power-law component
usually contributes a significant fraction of X-ray emission. 
Indeed, M101 ULX-1 showed a cool accretion disk spectrum 
($kT_{DBB}=0.16$ keV) during the December outburst.
Furthermore, 
the high state
luminosity is approximately
$16\%$ of the Eddington luminosity for a $6000\, M_\odot$ BH;
we therefore expect the inner disk to be optically thick, which is 
consistent
with the IMBH interpretation.

The state changes of \sss\ span a luminosity range larger than a factor
of $1000.$ \sss\
was first detected by \rosat\ in 1996 November \cite{wang99},
with an extrapolated unabsorbed 0.3--7 keV luminosity of $8\times10^{38}$ 
\lum\
(blackbody
model with $N_H=10^{21}$ cm$^{-2}$, $kT=75$ eV). Higher absorption
($3\times10^{21}$ cm$^{-2}$) would imply a luminosity up of
$4\times10^{39}$\lum. The source was not detected in other \rosat\
observations \cite{wang99}. In a 108 ks \rosat\ HRI
observation taken in 1996 May/June, we can set a
$3\sigma$ detection limit at $6.8\times10^{37}$\lum\ (power-law model
with $N_H=1.2\times10^{20}$ cm$^{-2}$ and $\alpha=2$). Setting the
$N_H$ at $1.5\times10^{21}$ cm$^{-2}$, the luminosity limit becomes
$1.5\times10^{38}$\lum. In 2000 March and October, \chandra\ detected
the SSS in the very soft state with luminosities of $\sim 10^{39}$\lum\ 
\cite{mukai03,dk03}.
In 2002 June, \xmm\ observed M101 and did not
detect the SSS \cite{jen04}. We re-analysed the image and found 
that the SSS
may be barely detected with possible contamination from nearby sources. We
derived the 0.3--7 keV luminosity of the SSS as
$\sim 10^{37}$\lum, by assuming a power-law model. The source was also in
the low state between 2004 January and May with luminosity of
$\sim10^{37}$\lum. It is clear that
the SSS has had at least 5 major outbursts ($L_X > 10^{39}$\lum). On 
the other hand, the
very low luminosities ($\sim 10^{37}$\lum)
 during 2002 and 2004 indicate that the source was in the low state.
The source varies by as much as a factor of 1000 between the
low state and the high state. This amplitude is even greater than many
Galactic BHs. Remarkably, the source also shows spectral changes.
It is very clear that there is a power-law
component ($\alpha=1.4$) in the composite low state spectrum, while the
high state spectra are supersoft. The \xmm\ observation taken
during the decay of the July outburst revealed that the blackbody
component was still strong and there was a very hard power-law
tail, similar to the low state.  In addition, the source showed spectral 
change during the 2004 December outburst. During the peak of the outburst, 
the X-ray spectrum changed from a supersoft spectrum to a quasisoft 
spectrum \cite{dk04}. It is clear that the spectrum was significantly 
harder than typical outburst supersoft spectra.  
This is the first and only ULX for which recurrent outbursts with state
transitions
similar to Galactic X-ray binaries have been observed.

With a peak bolometric luminosity $\approx 10^{41}$\lum, it is
likely that IMBH is the central engine of the system. Interestingly,
the blackbody component is always seen while the power-law component
becomes stronger when the source is at lower
luminosities. It may indicate that the power-law component is due to
Comptonization of soft photons \cite{wang04}. The photon index is,
however, harder
than that of Galactic BHs in the low state ($\alpha\sim1.7$). None of
the Galactic BHs has similar spectrum \cite{mr03}.
The closest example is Galactic microquasar V4641 Sgr for which the
photon index was measured to be between 0.6 and 1.3 \cite{miller02,mr03}.

Another important feature of our X-ray spectral fits is the presence
of absorption edges, which we found at 0.33,
0.57, 0.66, and 0.88 keV in two of the high state spectra. These are
consistent with C {\small V}, N {\small VI}, N {\small VII}, and O {\small 
VIII} edges. We note, however,
that the 0.33 keV edge may be due to calibration of the ACIS-S near
the carbon edge around 0.28 keV.
These features may signal the presence of highly ionized
gas in the vicinity of the accretor (e.g., warm absorber), and may be 
consistentwith an outflow from the source.
In fact, outflow models
have been suggested for ULXs in which the 
soft
component of some ULXs is due to outflow of a stellar-mass BH \cite{kp03}. 
Similar argument to explain the
super-Eddington luminosity of \sss\ \cite{mukai03}. However, the new data 
which reveal the 
changes of 
temperature and bolometric luminosity, the extremely high
luminosity, and the state transition of the source will be difficult
to explain by such a model. It
remains a puzzle that we saw different edges in the two observations,
possibly related to the geometry of the system.
Nevertheless, outflows from IMBHs may be expected.

It is also interesting to note that similar absorption edges
are expected and have been
observed in white dwarf (WDs) systems such as the recurrent nova
U Sco \cite{ka99}, CAL~87 and RX~J0925.7--4758 \cite{eb01}.
SSS radiation  with
luminosities in the range from $10^{37}$ ergs s$^{-1}$ to
roughly the Eddington limit for a $1.4\, M_\odot$ WD are
expected for nuclear burning WDs \cite{van92}. Indeed,
such models have been proposed to explain SSSs in the
Galaxy and Magellanic Clouds. Note, however, that neutron  star and
BH models have also been proposed. The high state luminosity of this 
source
is almost two orders of magnitude larger than that of any other
known SSS, and it rules out steady nuclear burning on hot WD models as an 
explanation for \sss.
A similar argument can be made for neutron star models.
A nova explosion could explain some features of the data. It is not 
favored, however, because the photospheric radius and lack of hard 
radiation at the peak of
the outburst have no obvious explanation.

We expect that at least some of the ULXs involving IMBH accretors are
transients \cite{king01,kal04}.
The required condition is that the donor must be a
massive star ($\gaeq$ 5 $M_{\odot}$) in regions of young populations. M101
ULX-1 satisfies these conditions. The colors of the
optical counterpart are consistent with an OB star. More recent optical 
observations suggest that the optical counterpart is a
B supergiant with a mass of
$9-12 M_{\odot}$ and the optical spectrum indicates that 
M101 ULX-1 is consistent with a high-mass X-ray binary \cite{kuntz05}. The 
source is also very close to star forming regions in a spiral 
arm as indicated in Figure 2.

It may be impossible to conclusively establish, using current technology, 
that
any X-ray source in an external galaxy is an IMBH.
A system like \sss\ is therefore particularly valuable, because
it provides $5$ different pieces of evidence that together
make a consistent argument in favor of an IMBH interpretation.
Its high-state luminosity, its short-time-scale
variability, its soft high-state spectrum, its pronounced
spectral changes, and the high-mass companion all suggest that \sss\ is a 
strong IMBH candidate.

\bigskip 
\begin{acknowledgments}
This work was supported by NASA grant NNG04GP58G.
A.K.H.K. acknowledges support from NASA grant GO3-4049X from the
Chandra X-Ray Center, and would like to
thank \xmm\ project scientist
Norbert Schartel for granting our DDT request.
We would also like to thank Matthew Jarvis (INT), Christian Veillet
(CFHT), Chris Benn (WHT), Richard Green (NOAO), and George Jacoby
(WIYN) for granting the target-of-opportunutiy time. 
This work is based on observations obtained with \xmm, an ESA mission with
instruments and contributions directly funded by ESA member states and
the US (NASA). The National Radio Astronomy Observatory is a facility 
of the National
Science Foundation operated under cooperative agreement by Associated
Universities, Inc.
\end{acknowledgments}

\bigskip 

\end{document}